\documentclass[fleqn,usenatbib]{mnras}

\usepackage{newtxtext,newtxmath}

\usepackage[T1]{fontenc}
\usepackage{ae,aecompl}

\usepackage{graphicx}	
\usepackage{amsmath}	
\usepackage{amssymb}	

\usepackage{xspace}

\newcommand{\AREPO}{\texttt{{AREPO}}\xspace}

\newcommand{\changaMM}{\texttt{{MANGA}}\xspace}
\newcommand{\changa}{{\texttt{ChaNGa}}\xspace}

\newcommand{\IllinoisGRMHD}{{\texttt{IllinoisGRMHD}}\xspace}
\newcommand{\NRPy}{{\texttt{NRPy+}}\xspace}

\newcommand{\SENRNRPy}{{\texttt{SENR/NRPy+}}\xspace}

\newcommand{\be}{\begin{eqnarray}}
\newcommand{\ee}{\end{eqnarray}}

\newcommand{\grad}{\ensuremath{\boldsymbol{\nabla}}}
\newcommand{\vel}{\ensuremath{\boldsymbol{v}}}

\newcommand{\ddt}[1]{\ensuremath{\frac{\partial #1}{\partial t}}}

\newcommand{\state}{\ensuremath{\boldsymbol{\mathcal{U}}}}
\newcommand{\charge}{\ensuremath{\boldsymbol{U}}}

\newcommand{\rhostar}{\ensuremath{\rho_*}}
\newcommand{\tautilde}{\ensuremath{\tilde{\tau}}}
\newcommand{\Svectilde}{\ensuremath{\tilde{\boldsymbol{S}}}}
\newcommand{\rtgamma}{\ensuremath{\sqrt{\gamma}}}
\newcommand{\T}[2]{\ensuremath{T^{#1 #2}}}
\newcommand{\uvec}{\ensuremath{\boldsymbol{u}}}

\newcommand{\vfluid}{\ensuremath{\boldsymbol{v}_{\rm f}}}

\newcommand{\flux}{\ensuremath{\boldsymbol{\mathcal{F}}}}
\newcommand{\fluxV}{\ensuremath{\boldsymbol{F}}}
\newcommand{\source}{\ensuremath{\boldsymbol{\mathcal{S}}}}
\newcommand{\sourceV}{\ensuremath{\boldsymbol{S}}}

\newcommand{\normal}{\ensuremath{\hat{\boldsymbol{n}}}}

\newcommand{\meshv}{\ensuremath{\boldsymbol{w}}}
\newcommand{\facev}{\ensuremath{\boldsymbol{\tilde{w}}_{ij}}}
\newcommand{\facer}{\ensuremath{\boldsymbol{\tilde{r}}_{ij}}}
\newcommand{\meshr}{\ensuremath{\boldsymbol{r}}}

\newcommand{\Tmunu}{\ensuremath{T^{\mu\nu}}\xspace}

\title{General Relativistic Hydrodynamics on a Moving-mesh I: Static Spacetimes}

\author[Chang \& Etienne]{Philip Chang${}^{1}$ and Zachariah B. Etienne${}^{2,3}$\\
$^1$ Department of Physics, University of Wisconsin-Milwaukee, 3135 North Maryland Avenue, Milwaukee, WI 53211; chang65@uwm.edu\\
$^2$ Department of Physics and Astronomy, West Virginia University, Morgantown, WV 26506; zbetienne@mail.wvu.edu\\
$^3$ Center for Gravitational Waves and Cosmology, West Virginia University, Chestnut Ridge Research Building, Morgantown, WV 26505}

\date{Accepted XXX. Received YYY; in original form ZZZ}

\pubyear{2015}

\begin{document}
\label{firstpage}
\pagerange{\pageref{firstpage}--\pageref{lastpage}}
\maketitle

\begin{abstract}
We present the 
moving-mesh general relativistic hydrodynamics solver for static
spacetimes as implemented in the code, \changaMM. Our implementation builds on the
architectures of \changaMM and the numerical relativity Python package \NRPy. We review the
general algorithm to solve these equations and, in particular,
detail the time stepping; Riemann solution across moving faces; 
conversion between primitive and conservative variables; 
validation and correction of hydrodynamic variables; and mapping of
the metric to a Voronoi moving-mesh grid.
We present test results for the numerical integration of an
unmagnetized Tolman-Oppenheimer-Volkoff star for 24 dynamical times.
We demonstrate that at a resolution of $10^6$ mesh generating points,
the star is stable and its central density drifts downward by 2\% over
this timescale. At a lower resolution the central density drift
increases in a manner consistent with the adopted second order spatial
reconstruction scheme. These results agree well with the exact
solutions, and we find the error behavior to be similar to Eulerian 
codes with second-order spatial reconstruction. We also demonstrate
that the new code recovers the fundamental mode frequency for the same
TOV star but with its initial pressure depleted by 10\%.
\end{abstract}




\section{Introduction}

Multimessenger observations are beginning to address some of the
most important unsolved problems in gravity and astrophysics. These
include testing strong field general relativity (GR), constraining the
nuclear equation of state (EOS), elucidating the origin of the elements, and
uncovering plausible formation scenarios for black hole and neutron star
binaries.  However, this promise is predicated on having detailed
theoretical models of multimessenger sources, in particular of compact
binary merger events.

Construction of self-consistent theoretical models remains a central problem
in the theory of compact object mergers, and their complexity cannot be
overstated.  There are several reasons for this.  First, the fluid
equations in dynamical spacetimes are considerably more complex than
in the Newtonian limit; both the Einstein field equations
and the those of general relativistic (magneto)hydrodynamics
(GR(M)HD) must be solved. Second, the equations must be evolved over many orders of
magnitude in length and time, spanning from the core of the
neutron star (NS) out to the far-field wave zone.
Third, this is a multiphysics problem, spanning the range from
hyperaccretion physics at a black hole horizon to the core of a
neutron star to interstellar space and includes the physics of
radiation, neutrino radiation, nuclear reactions, nuclear equations of
state, gravitational waves, shocks, and accretion, just to name a few.

Fully three-dimensional numerical (magneto-)hydrodynamic simulations are required to accurately
model and understand these systems. 
In the context of binary neutron star GR(M)HD simulations, for
example, a number of numerical codes have been developed to solve the
GRMHD equations. Most codes solve the Einstein field
equations without approximation either 
on an adaptive-mesh refined (AMR) grid or (pseudo-)spectrally and solve the equations
of GR(M)HD on an AMR grid
(e.g., \citealt{Duezetal2005,Anderson:2006ay,2007CQGra..24S.235G,CerdaDuran2008,2008PhRvD..78f4054Y,2008PhRvD..78j4015D,Dionysopoulou:2012zv,Palenzuela2013,GRHYDRO,IllinoisGRMHD,SpECTRE}).
Others solve the Einstein
equations in the conformally-flat approximation and solve the
equations of GRHD with smoothed-particle-hydrodynamics (SPH)
methods \citep{2002PhRvD..65j3005O}. 

These two methodologies for solving the (M)HD equations---smooth
particle hydrodynamics (SPH) and grid-based 
solvers---have their respective advantages and
disadvantages\footnote{As this paper focuses on a new GRHD solver for
  {\it static} spacetimes, we do not contrast the dynamical GR
  solvers. However, future plans do include implementation of an
  Einstein field equation solver (without approximation) based on
  \SENRNRPy, as described in the Conclusions and Future Work section.}.
Smooth particle hydrodynamics (SPH) is based upon the Lagrangian view
of the Euler equations where the sampling of a fluid is determined
from a finite number of particles, and fluid quantities like density
and pressure are determined by computing a smoothing kernel over a
number of neighbors.  The Lagrangian nature of SPH allows it to
conserve linear and angular momentum, but comes at the expense of
comparatively poor resolution of shocks due to its smoothing nature.
On the other hand, grid based methods have superior shock capturing
abilities due to the use of Godonov schemes, but suffer from grid
effects, e.g., the presence of grid direction can affect the
conservation of angular momentum. 

\citet{2010MNRAS.401..791S} developed an arbitrary Lagrangian-Eulerian
(ALE)/moving-mesh (MM) scheme in an effort to capture the best characteristics
of both approaches. This scheme, which is implemented in \AREPO \citep{2010MNRAS.401..791S,2019arXiv190904667W}, relies on a Voronoi tessellation to generate well-defined and unique meshes for an arbitrary distribution of points that deform continuously under the movement of the mesh generating points. 
\citet{2010MNRAS.401..791S} has argued that the use of ALE schemes is important in maintaining
the Galilean invariance of Eulerian schemes in the Newtonian case.  It has also been argued
that these schemes are superior to SPH
and Eulerian grid schemes at capturing boundary layer
instabilities such as Kelvin-Helmholtz instabilities  (\citealt{2010MNRAS.401..791S} but also
see~\citealt{2016mnras.455.4274l}). In any case they do seem
ideal for modeling colliding galaxies or stars. 
In particular, \AREPO, has been used to study a number of different astrophysical problems including cosmological galaxy formation \citep[see for instance][]{2014MNRAS.444.1518V}, disks, and stellar mergers \citep{2015ApJ...806L...1Z,2016ApJ...816L...9O}.

Aside from \AREPO, a number of MM codes have been developed based on this scheme.  These include \texttt{TESS} \citep{2011ApJS..197...15D}, \texttt{FVMHD3D} \citep{2012ApJ...758..103G}, \texttt{ShadowFax} \citep{2016A&C....16..109V}, \texttt{RICH} \citep{2015ApJS..216...35Y}, \texttt{DISCO} \citep{2016ApJS..226....2D}, and \changaMM \citep{Chang+17}.
These MM schemes have also been extended to include magnetic fields
(\citealt{2011MNRAS.418.1392P,2014MNRAS.442...43M,2016MNRAS.463..477M}),
higher-order convergence \citep{2016MNRAS.455.1134P,2015MNRAS.452.3853M},
and new physics, such as cosmic rays
\citep{2016MNRAS.462.2603P,2017MNRAS.465.4500P}.  In addition, the
general scheme of determining the geometry from an arbitrary
collection of points has also led to derivative methods such as
\texttt{GIZMO} \citep{2015MNRAS.450...53H}.

We have recently developed \changaMM, a MM hydrodynamic solver for \changa \citep{Chang+17}, which is largely based on the \citet{2010MNRAS.401..791S} scheme.  \changaMM is geared toward the study of dynamical stellar problems, such as common envelope evolution \citep{Prust+19, Prust20} and tidal disruption events (Spaulding \& Chang, in preparation).  We have also been steadily adding new physics including radiation hydrodynamics \citep{Chang+20}, magnetic fields, and various equations of state (EOSs) such as the \texttt{HELMHOLTZ} EOS \citep{2000ApJS..126..501T}, the \texttt{MESA} EOS \citep{2011ApJS..192....3P,2013ApJS..208....4P,2015ApJS..220...15P}, and a nuclear EOS \citep{2010CQGra..27k4103O,2017PhRvC..96f5802S}.  In addition, we have also developed moving, reactive boundary conditions \citep{Prust20} and a multi-stepping scheme \citep{Prust+19}. 

The algorithmic advantages of the ALE scheme in its ability to capture
both shocks and interface instabilities make it exceptionally
well-suited for an application to
GR(M)HD for the problem of compact object mergers.  In this paper, we describe ongoing work toward this goal.  Here, we describe our extensions to \changaMM which enables it to solve the equations of GRHD on \textit{static} spacetimes.  Future work will tackle the problem of dynamical spacetimes. We note that this work is similar to \citet{2017ApJ...835..199R}, who implemented GRHD in \texttt{DISCO}.  The difference is that the work described here is for a completely arbitary unstructured moving mesh.

This paper is organized as follows.  We write the equations of GRHD in
a flux-conservative form that can be solved on a moving-mesh and
pictorially describe such a scheme in \S \ref{sec:hydro}.  We provide
an overview of the algorithmic steps in \S \ref{sec:numerical} and highlight specific
technical details such as the conservative-to-primitive solver, the
time integrations, the Riemann solver, the metric map, and the
variable validation.  We describe a number of test problems in \S
\ref{sec:test problems} including a Tolman-Oppenheimer-Volkoff
(TOV)~\citep{1934rtc..book.....T,1939PhRv...55..374O} star and stellar
oscillations.  We conclude in \S \ref{sec:conclusions} and close with
a discussion of the road forward.

\section{General Relativistic Hydrodynamics}\label{sec:hydro}

In the following equations we adopt $G=c=1$ units and the Einstein
summation convention. For expressions involving tensors, Latin indices
denote the spatial components and Greek indices space {\it and}
time components.

The equation of hydrodynamics in arbitrary spacetimes can be written
in conservative form (see for instance \citealt{Duezetal2005}, who
adopt the same formulation and variable conventions) by
introducing a state vector $\state=(\rhostar, \Svectilde, \tautilde)$:
\be
\ddt{\state} + \grad\cdot\flux dV = \source\label{eq:state},
\ee
where $\rhostar = \alpha\rho\rtgamma u^0$, $\Svectilde = \rhostar h \uvec$, $\tautilde = \alpha^2\rtgamma \T00 - \rhostar$.
The flux, $\flux$ is given by
\be
 \flux=\left( \begin{array}{c} 
   \rhostar v^j\\ 
   \alpha\rtgamma\T{j}{\beta}g_{\beta i}\\
   \alpha^2\rtgamma\T0j - \rhostar v^j
 \end{array}\right)
\ee
where $v^j$ are the components of the 3-velocity ($\vel$ is its vector form). 
The source, $\source$, is
\be
\source = \left(\begin{array}{c}
   0\\ 
   \frac 1 2 \alpha\rtgamma \T{\alpha}{\beta}g_{\alpha\beta,i}\\
   s
\end{array}\right),
\ee
where 
\be
s=\alpha\rtgamma\left[\left(T^{00}\beta^i\beta^j + 2 T^{0i}\beta^j + T^{ij}\right)K_{ij} - \left(T^{00}\beta^i + T^{0i}\right)\partial_i\alpha\right].
\ee
The stress energy tensor for a perfect fluid is 
\be
\T{\alpha}{\beta} = \rho h u^{\alpha} u^{\beta} + P g^{\alpha\beta},
\ee
where $h = 1 + \epsilon + P/\rho$ is the specific enthalpy and
$u^{\alpha}$ are the respective components of the four velocity.

The associated set of primitive variables are $(\rho, \vel,
\epsilon)$, which are the rest mass density, fluid 3-velocity
$u^i/u^0$, and internal energy (measured in the rest frame).
We close the set of equations with a simple $\Gamma$-law EOS: $P =
(\Gamma - 1)\rho\epsilon$.  In this work we pick the adiabatic index,
$\Gamma$, to be equal to the polytropic index that we select for the
neutron star ($\Gamma=2$) discussed in \S~\ref{sec:test problems}. In doing so, we set the initial internal energy, $\epsilon$, using the polytopic EOS. 


\citet{2010MNRAS.401..791S} showed that any generic flux-conservative
equation (\ref{eq:state}) can be solved using a finite volume strategy
on a moving unstructured mesh.  For instance, the MHD equations can
also be cast in this form
\citep{2011MNRAS.418.1392P,2011ApJS..197...15D,2012ApJ...758..103G,2014MNRAS.442...43M,2016MNRAS.463..477M}.
We refer the interested reader to \citet{Chang+17}, \citet{Prust+19}, and
 \citet{Chang+20} for a more detailed
discussion of the scheme in \changaMM.  Here, we provide a
summary:

For each cell, the integral over the volume of the $i$th cell defines
the charge
of the $i$th cell, $\charge_i$, to be
\be
\charge_i = \int_i \state dV = \state_i V_i,
\ee
where $V_i$ is the volume of the cell. We then use Gauss' theorem to convert the volume integral over the divergence of the flux in equation (\ref{eq:state}) to a surface integral:
\be
\int_i \grad\cdot\flux dV = \int_i \flux\cdot\normal dA
\ee
We now take advantage of the fact that the volumes are Voronoi cells with a finite number of neighbors to define a integrated flux
\be
\sum_{j \in {\rm neighbors}} \fluxV_{ij} A_{ij} = \int_i \flux\cdot\normal dA,
\ee
where $\fluxV_{ij}$ and $A_{ij}$ are the average flux and area of the common face between cells $i$ and $j$.
The discrete time evolution of the charges in the system is given by:
\be
\charge_i^{n+1} = \charge_i^n + \Delta t \sum_j \hat{\fluxV}^{n+1/2}_{ij} A^{n+1/2}_{ij} + \Delta t\sourceV^{n+1/2}_i, \label{eq:time evolution}
\ee
where $\hat{\fluxV}^{n+1/2}_{ij}$ is an estimate of the half-timestep flux between the initial, $\charge_i^n$, and final states $\charge^{n+1}_i$; $A^{n+1/2}_{ij}$ is the time-averaged area of the face between $i$ and $j$;  and $\sourceV_i^{n+1/2} = \int_i \source dV$ is the time-averaged integrated source function. \footnote{We note that $\hat{\fluxV}^{n+1/2}_{ij}$ given by the Riemann flux solved in the ``rest'' frame of the face and boosted back into the ``lab'' frame.}

\section{Numerical implementation}\label{sec:numerical}

Because the GR(M)HD equations can be written in flux-conservative form
as well, they can also be solved on a moving unstructured mesh.  Our
algorithm is outlined below: 
\begin{enumerate}
  \itemsep0em
  \item At the beginning of a timestep, the Voronoi cells are built and the volume integrated charges, \charge, are mapped to the state vector, \state. 
  \item A predictor step is applied to obtain the half-timestep state
    vector. The conservative variables in the state vector are
    mapped to half-timestep primitive variables, via the
    conservatives-to-primitives solver.  \label{predictor}
  \item The mesh generating point is drifted a half-timestep forward and the Voronoi mesh is rebuilt on this half-timestep. 
  \item The primitive variables on the faces are reconstructed via slope-limited linear interpolation. This is combined with the metric to produce the state vector on the faces. 
  \item The state vector flux across the moving faces is estimated at
    the half-timestep, using the relativistic version of the
    Harten-Lax-van Leer (HLL) approximate Riemann
    solver \citep{HartenEtAl_1983}. \label{riemann}
  \item The cells are drifted another half timestep.
  \item All the fluxes are summed and the source terms are included to update the \charge of the cell to the full timestep. 
  \item The updated charges are mapped to a state vector, and the
    conservative variables in the state vector are mapped to the
    primitive variables (via the conservatives-to-primitives
    solver). The physicality of the primitives is checked and
    prescribed fixes are applied as needed. This marks the end of a
    timestep; return to the top as needed until the chosen final time
    is reached.
\end{enumerate}
We can also use the multi-stepping scheme as described in
\citet{Prust+19}, which we do by default. Here the key difference is
that each cell is associated with the largest timestep possible for
that cell from a factor-of-two hierarchy.  Each face is then
integrated on the smallest timestep of two neighbors that define it.
The changes to the charge of each cell over the timestep are
accumulated and then applied at the end of the cell's timestep.

Most of these steps are fairly straightforward moving-mesh methods.
We discuss below a few important technical details of the algorithm
that are significantly different than standard moving-mesh techniques
described in the literature, including the conservative-to-primitive
variable solver, details of the predictor-corrector time stepping, the
solution of the Riemann problem on moving faces in GRHD, the
validation of the primitive variables to ensure physically relevant
states, and the mapping of the metric to arbitrary points in
coordinate space.

The GRHD flux and source terms themselves are most cleanly written in
Einstein notation. Expanding these equations in full by hand in, e.g.,
the C language directly would be both time consuming and error
prone. \NRPy\footnote{\url{http://nrpyplus.net}}~\citep{Ruchlin2018}, ``Python-based code generation for Numerical
Relativity... and Beyond!'', converts equations written in Einstein
notation into highly optimized C-code kernels. The GRHD flux and source
terms, as well as equations needed for the primitives-to-conservatives
variable conversion, were written within the \NRPy
framework as part of this work.
The GRHD flux and source terms were validated against the hand-coded
implementations within \IllinoisGRMHD, and the primitive to/from
conservative equations were validated by converting many
physically valid sets of primitive variables to their
conservative form and back (using the Newton-Raphson-based root-finder
of \citealt{2006ApJ...641..626N} for the conservatives-to-primitives
step, as described below). The detailed Jupyter notebook used to
generate needed GRHD equations in \changaMM may be found 
in the \NRPy \texttt{github} repo\footnote{\url{https://github.com/zachetienne/nrpytutorial}} or be viewed
directly via \texttt{nbviewer}\footnote{\url{https://nbviewer.jupyter.org/github/zachetienne/nrpytutorial/blob/master/Tutorial-GRHD_Equations-Cartesian-c-code.ipynb}}.

Before proceeding, we note two important points.  
Previously, \changaMM used only cgs units, but for GR, $G=c=1$ units
are superior so the code was extended to support them. Second,
although our moving mesh code is limited to periodic boundary
conditions, no boundary conditions are applied to
the metric (as it is static and known) so there is no ``periodic
gravity.'' We simply ensure that the outer boundaries of our numerical 
domain are sufficiently far away from regions of interest, so that as
outer (periodic) boundary conditions are applied to the {\it hydrodynamic}
variables, they have little to no effect on the simulation.

In future work, we may as needed adopt radiation inflow and outflow boundary
conditions as described in \citet{Chang+20} to the hydrodynamic
variables on the boundary, to produce arbitrary boundary
conditions. We note that this implementation would be similar to the
boundary conditions for a ``sphered cube''
\citep{2019arXiv190312642B}. 


\subsection{Time Stepping}

We time integrate equation (\ref{eq:time evolution}) as follows:
\begin{enumerate}
   \item Estimate the Courant-limited timestep $\Delta t$ for each cell, as
     described in \citet{Chang+17}.  The timestep can either be an individual timestep in a multi-step algorithm \citep{Prust+19} or a global timestep \citep{Chang+17}.
   \item Estimate the half timestep state of the cell using the
     second-order van Leer scheme described in \citet{Chang+20}. In brief, we solve the RHS of equation (\ref{eq:time evolution}), but with the replacement of $\Delta t \rightarrow \Delta t/2$ and use piecewise continuous reconstruction to get face values.  We then solve for the fluxes following the methodology described below and include the source terms on a half-timestep.
   \item Drift the mesh generating points by a half-timestep and
     recompute the half-timestep tessellation to provide second order
     convergence in time (e.g., provide an estimate for $A^{n+1/2}_{ij}$).
   \item Use the half-timestep states and use full linear reconstruction to derive face values to compute the half-timestep fluxes, $\hat{\fluxV}^{n+1/2}_{ij}$ (described below). Then apply the full step including the source terms (using the half-step values), e.g., $\sourceV_i^{n+1/2}$.
   \item Update the state of the cell to the full step. 
\end{enumerate}
As stated in \citet{Chang+20}, the use of the van Leer half-step prediction allows the source terms to be automatically included at second order. It also simplifies the code as the equations are only written once.  Moreover, the van Leer method can be easily adapted to multi-stepping schemes as we have done here.  The only change is that the half-timestep estimate must be taken from the cell's initial state. 

\changaMM was originally designed to reconstruct the conserved
values at cell faces.  For Newtonian codes, this is a reasonable
choice, but modeling highly relativistic flows common in GRHD
requires reconstructing the primitive variables instead.  This is the
scheme that we have now implemented in \changaMM. The reconstruction of
these primitive variables follows the linear reconstruction scheme
described in \citet{Chang+17}. After reconstruction, we validate our
face primitive values as outlined in \S~\ref{sec:validation}.  We note
that an even better scheme would be to reconstruct the modified state
vector $(\rho, \uvec, \epsilon)$, e.g., replacing the fluid 3-velocity
with the 3 component of the four-velocity \citep{1999ApJS..122..151A,2013A&A...560A..79L}.  Reconstruction along this choice
guarantees that the reconstructed $\vfluid$ and $\vel$ are
valid (though one still must be careful to ensure Lorentz factors do
not get too large).

\subsection{Riemann Solution across Moving Faces}

The half-timestep flux across each face, $\hat{\fluxV}^{n+1/2}_{ij}$,
is estimated using an approximate Riemann solver.  Previously, the 1-D
fluxes are computed across each face in the \textit{Galilean} rest
frame of that face and then collectively applied each timestep
\citep{Chang+17}.  In short, the steps involved are:
\begin{enumerate}
 \item Estimate the velocity $\facev$ of the face \citep{2010MNRAS.401..791S,Chang+17}:
 \be\label{eq:face velocity}
 \facev = \frac{(\meshv_i - \meshv_j)\cdot(\facer - 0.5(\meshr_j+\meshr_i))}{|\meshr_j - \meshr_i|}\frac{\meshr_j - \meshr_i}{|\meshr_j - \meshr_i|} + \bar{\boldsymbol{w}}_{ij},
 \ee
 where $\bar{\boldsymbol{w}}_{ij} = 0.5(\meshv_i + \meshv_j)$ is the average velocity of the two mesh generating points and \facer\ is the face center between cells i and j.
\item Estimate the half-timestep state vector (in the rest frame of the moving face) at the face center (\facer) between the neighboring $i$ and $j$ cells via linear reconstruction. \label{item:half step}
 \item Boost the state vector from the ``lab'' frame to the rest frame of the face center and rotate the state vector such that the x-axis points along the outward normal of the face, i.e., in the direction from $i$ to $j$.  Note that these boosts are Galilean in the coordinate space \citep[See in particular][]{2011ApJS..197...15D} 
 \item Estimate the flux using a 1-D Riemann solver.    
 \item Boost the solved flux back into the ``lab'' frame.
\end{enumerate}
These steps are clear in a Newtonian contexts as the Galilean boosts involve only upper index vectors.  However, the action of changing reference frames is far less trivial in an arbitrary spacetime geometry.  Thus, we are forced to return to the basic equations to derive a new (simpler) scheme.  To do this, consider the integral of the flux over a Voronoi cell:
\be
\int_i \flux\cdot\normal dA = \sum_{j \in {\rm neighbors}} \hat{\fluxV}^{n+1/2}_{ij}\cdot\normal_{ij} A^{n+1/2}_{ij}
\ee
Taking the direction of $\normal_{ij}$ to denote the ``left'' and ``right'' states of a face, we can write the ``left'' and ``right'' states of a face to be $\state_L$ and $\state_R$.  Following the methodology of \citet{2011ApJS..197...15D}, we solve the Riemann flux and state in the ``lab'' frame.  The flux across the moving face is then $\flux' = \flux_{\rm Riemann} + \facev\cdot\normal\state_{\rm Riemann}$., where $\flux_{\rm Riemann}$ and $\state_{\rm Riemann}$ are the Riemann flux and state in the ``lab'' frame, respectively.
Written in this way, we avoid the issues with boosting to a
face-oriented coordinate system and the overall code is simpler. 

We are finally left with the choice of the Riemann solver.  For our
purposes, the relativistic generalization of the HLL solver
\citep{HartenEtAl_1983,toro2009riemann,Duezetal2005}
\be
\fluxV_{\rm HLL} = \frac{\lambda^+\fluxV_L -\lambda^-\fluxV_R + (\state_R-\state_L)\lambda^+\lambda^-}{\lambda^+-\lambda^-}
\ee
is adopted, where the wave speeds are defined as 
\be
\lambda^+ = \max(\lambda^+_R,\lambda^+_L) \quad\textrm{and}\quad \lambda^- = \min((\lambda^-_R,\lambda^-_L)).
\ee
The estimated L,R wave speeds are given by the quadratic equation: 
\be
a_1(\lambda^{\pm}_{L,R})^2 +a_2\lambda^{\pm}_{L,R} +a_3 = 0,
\ee  
where the $a_i$'s are given by
\be
a_1 &=& (1-c_s^2)(u^0)^2 - c_s^2g^{00},\\
a_2 &=& 2c_s^2g^{\hat{n}0}- 2u^iu^0(1-c_s^2),\\
a_3 &=& (1-c_s^2)(\uvec\cdot\normal_{ij})^2 - c_s^2 g^{\hat{n}\hat{n}}.
\ee
Here, the upper index of $\hat{n}$ denotes the component along
the face's normal direction. Note that in the case of MHD, there is a
replacement of $c_s^2 \rightarrow v_0^2 = v_A^2 + c_s^2(1-v_A^2)$,
where $v_A$ is the Alfv\'en velocity. We will note that the
  \citet{2011ApJS..197...15D} uses an HLLC solver instead of the HLL
  solver used here, which has the advantage that when the cells are
  moving close to the fluid velocity, the advective flux is nearly
  exactly canceled by the face velocity term. Thus the HLLC solver has
  the advantage of preserving contact discontinuties.

\subsection{Conversion and Validation of Hydrodynamic Quantities}\label{sec:validation}

All GRHD quantities can be constructed from the primitive variables,
which include density $\rho$, velocity $\vel$, and internal energy
$\epsilon$.  The conversion to the conservative variables is
algebraically straightforward provided the local metric quantities are
known.  However, the conversion from the conservative to primitive
variables cannot generally be accomplished in GRHD by simple algebraic means
and must be solved using an algebraic or numerical root-finding algorithm. 
Here we use the publicly available conservatives-to-primitives solver by
\citet{2006ApJ...641..626N}.

Prior to each conservatives-to-variables conversion, conserved
variables are checked so that they are physically valid.  The
checks that we employ are $\rhostar > 0$ and $\tautilde > 0$. Cells
that violate these physical checks have their primitive variables
reset above the floor and new conserved variables are calculated. For
such cells this represents the entirety conservatives-to-primitives
conversion process.

For cells with physically valid conservative variables, the 
\citet{2006ApJ...641..626N} conservatives-to-primitives solver is
called. Immediately after the conservatives-to-primitives conversion,
we reset the density if it falls beneath a value, $\rho_{\rm min}$
(i.e., we impose a low-density atmosphere).
We also ensure that the velocities do not become unphysically large by
capping Lorentz factor to a maximum, $\Gamma_{\rm max}$.

We expand on this point by considering the Valencia 3-velocity $v^i_{(n)}$: 
\be
\alpha v^i_{(n)} = v^i + \beta^i = \frac{u^i}{u^0} + \beta^i
\ee
where $v^i = \frac{u^i}{u^0}$ and $u^{\mu}$ is the four velocity. The Lorentz factor in this case is: 
\be
\Gamma = \sqrt{\frac{1}{1 - \gamma_{ij}v^i_{(n)}v^j_{(n)}}}
\ee

Numerical errors especially near large density gradients will
occasionally drive the denominator in the radical to negative
or very tiny values. Thus, to ensure that the values remain physical,
we limit the potentially offending term to be: 
\be
\gamma_{ij}v^i_{(n)}v^j_{(n)} = 1 - \Gamma^{-2} < 1-\Gamma_{\rm max}^{-2}
\ee
and adjust the corresponding 3-velocity when the above condition is violated to be:
\be
v^i_{(n)} = \sqrt{\frac{1 - \Gamma_{\rm max}^{-2}}{1-\Gamma^{-2}}}v^i_{(n)}.
\ee

\subsection{Metric Quantities}

In this paper, we assume only static spacetime metrics.  Many static
metrics of great astrophysical interest have closed-form expressions,
e.g., Kerr and Schwarzschild.  However, in this paper we
will focus on the TOV metric, which does not possess such a solution.
However, as the TOV metric is spherically
symmetric and static, it can be computed from the solution of a set of
simple ordinary differential equations on a radial grid at extremely high resolution.  For the
TOV metric, the Arnowitt-Deser-Misner (ADM) 3+1 line element can be represented by:
\be\label{eq:adm metric}
ds^2 = -\alpha^2 dt^2 + \gamma_{rr} dr^2 + \gamma_{\theta\theta} d\theta^2+ \gamma_{\phi\phi} d\phi^2,
\ee
where $\gamma_{\theta\theta} = r^2$, and $\gamma_{\phi\phi} = r^2
\sin^2 \theta$. The other two functions $\alpha$ and $\gamma_{rr}$ can
be interpolated from the TOV solution on the dense radial grid. As all
our evolutions are in the Cartesian basis, we must in addition perform
the necessary spherical-to-Cartesian basis transformation for each
tensor.

\section{TOV Star Code Tests}\label{sec:test problems}

We validate our GRHD extensions to \changaMM with a set of two very
challenging code tests {\it in full 3D}. In the first test, we evolve
TOV initial data (\S~\ref{TOVID}) with a fixed background spacetime
(\S~\ref{TOVevol}). This is a useful code test, as the exact solution
is stationary; thus any dynamics in time are purely a result of
numerical errors. We apply this fact to directly measure the rate at which
our numerical errors converge to zero with increased numerical resolution.
In the second test, we evolve the same initial data but with the
initial pressure in the TOV star depleted by 10\% (\S~\ref{TOVoscilevol}),
and compare the oscillations induced with those observed in a trusted
code \IllinoisGRMHD at very high resolution.

We performed these computations on the Stampede 2 supercomputer at the Texas Advanced Computing Center, the Niagara supercomputer at the University of Toronto \citep{Loken_2010,10.1145/3332186.3332195}, and Thorny Flat HPC System at West
Virginia University.

\subsection{TOV Initial Data}
\label{TOVID}

The TOV equations are 
\begin{eqnarray}
\frac{dP}{dr} &=& -\mu\frac{M}{r^2}\left(1 + \frac P {\mu }\right)\left(1 + \frac {4\pi r^3 P}{M}\right)\left(1 - \frac {2M}{r}\right)^{-1}\\
\frac{dM}{dr} &=& 4\pi \mu r^2.
\end{eqnarray}
The system of equations is closed by choosing a
polytropic EOS $P = \rho^2$ (consistent with cold,
degenerate nuclear matter).  In setting up the initial conditions with
\changaMM's $\Gamma$-law EOS ($P=(\Gamma-1)\rho \epsilon$), we simply
set $\epsilon = \rho$ and $\Gamma=2$.

The $M$ above is the rest mass measured outside the star (i.e., at
$r>R$). Note this is different from the mass measured by integrating
the mass-energy density $\mu = \rho h$ over the proper volume
\begin{equation}
M' = \int_0^{\infty} \frac{4\pi r^2\mu}{\sqrt{1 - \frac {2 M}{r}}} dr
\end{equation}
We note (with caution) that much of current literature uses $\rho$ to
denote the mass-energy density $\mu$, which can be potentially confusing.

We numerically solve these ordinary differential equations to set up TOV
initial data for our simulations, as follows.  First we pick a central baryonic
mass density $\rho_{0,c} = 0.129285$, then we compute a central
pressure $P_c$ and central mass-energy density $\mu_c$.  At $r=0$, we
assume that $\mu=\mu_c$ is a constant and numerically integrate
outward until the pressure is $10^{-8}$ of the central pressure.
\begin{figure*}
   \includegraphics[width=0.3\textwidth]{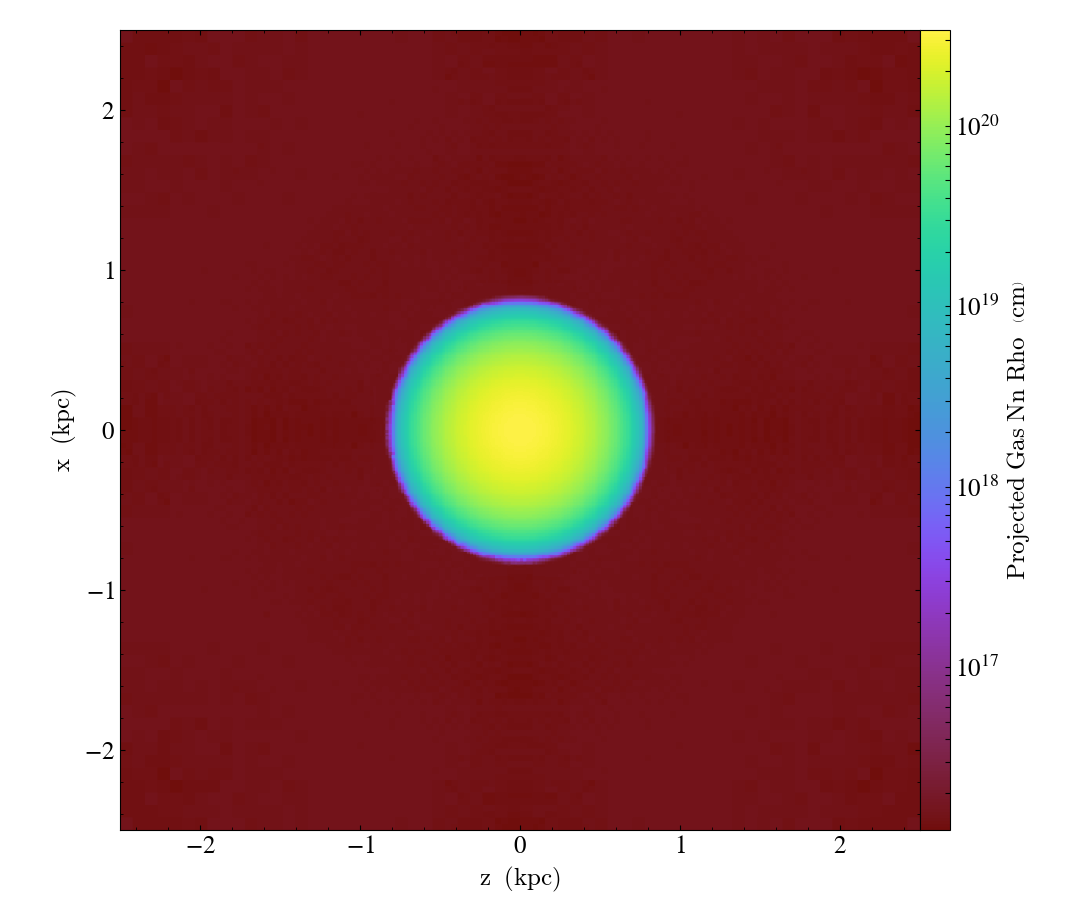}   \includegraphics[width=0.3\textwidth]{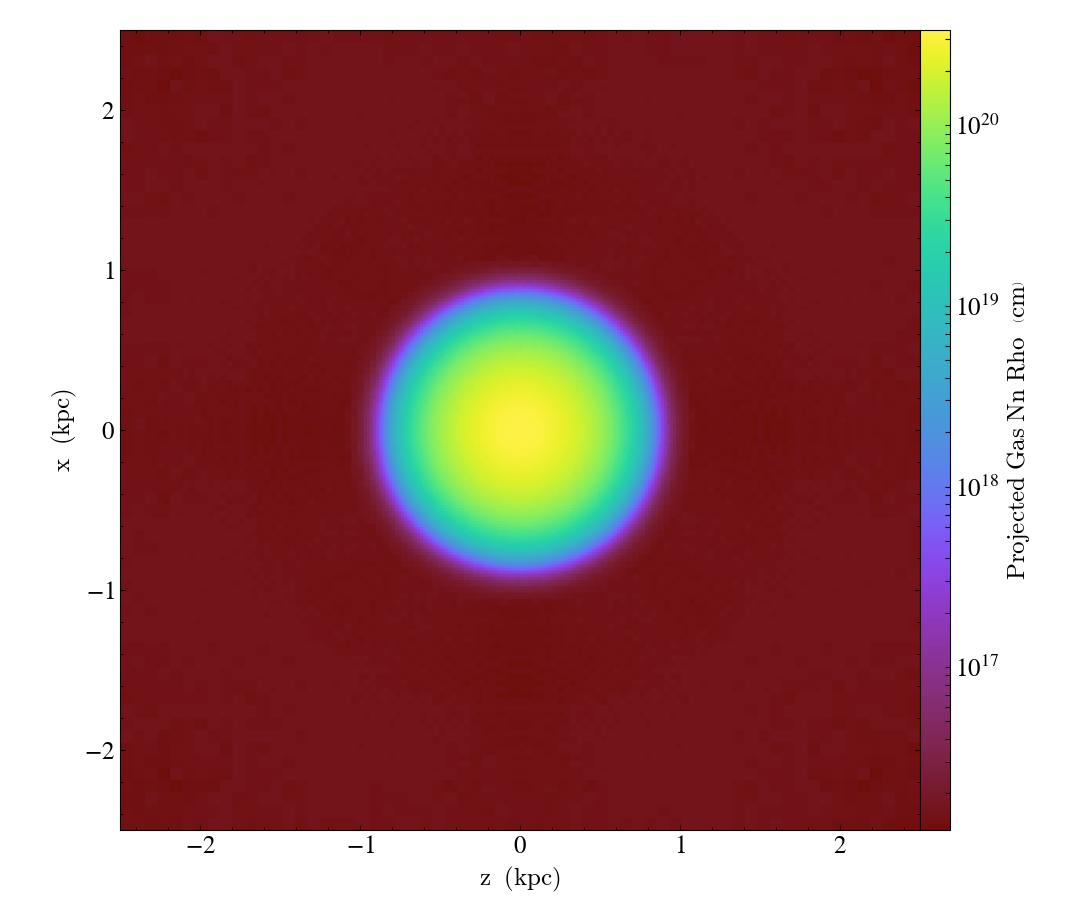}
   \includegraphics[width=0.3\textwidth]{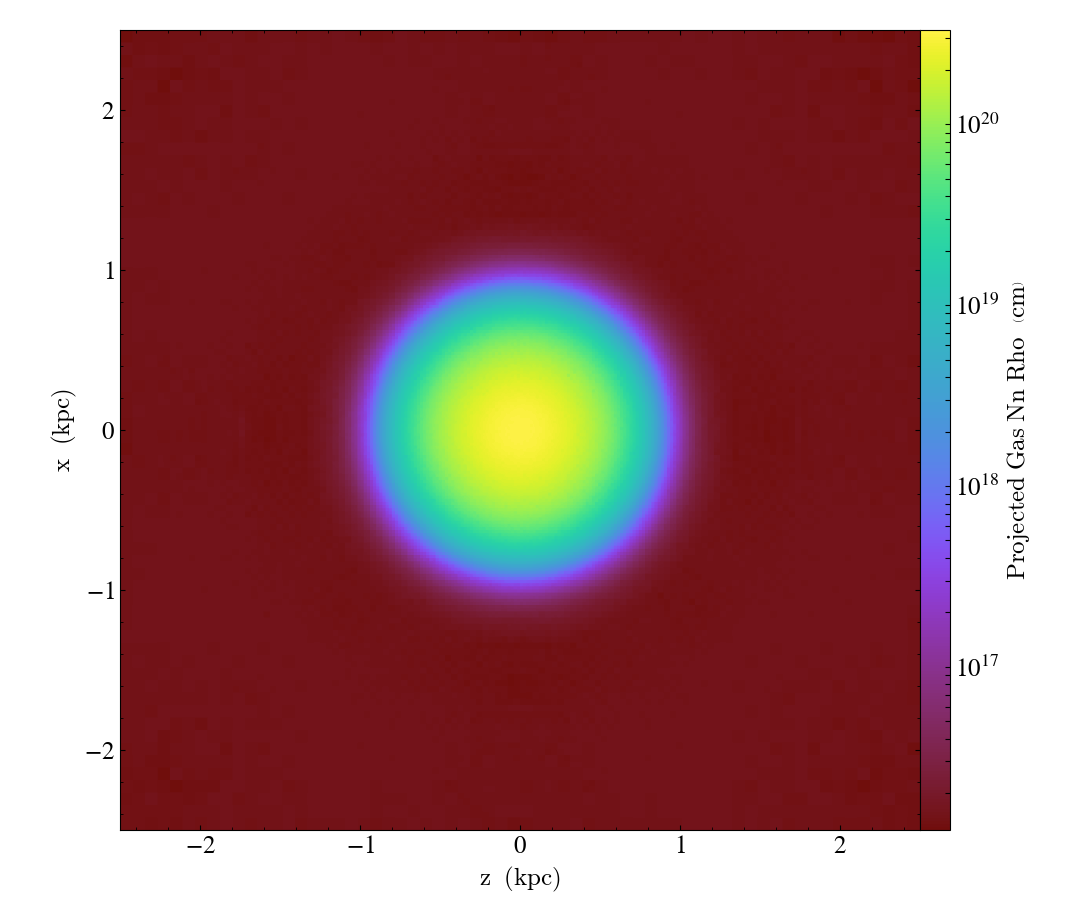}\\
   \includegraphics[width=0.3\textwidth]{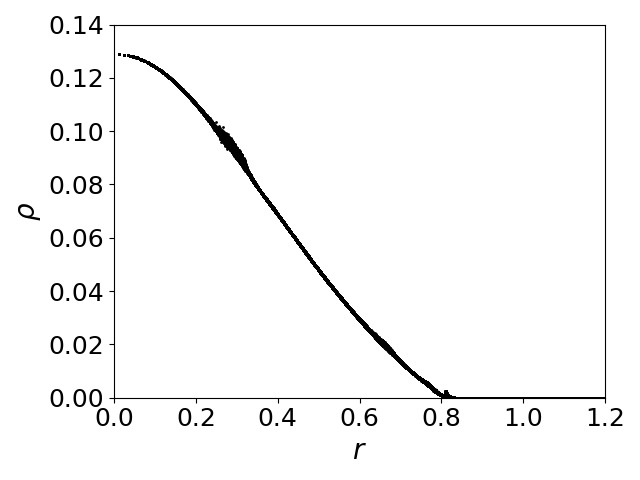}   \includegraphics[width=0.3\textwidth]{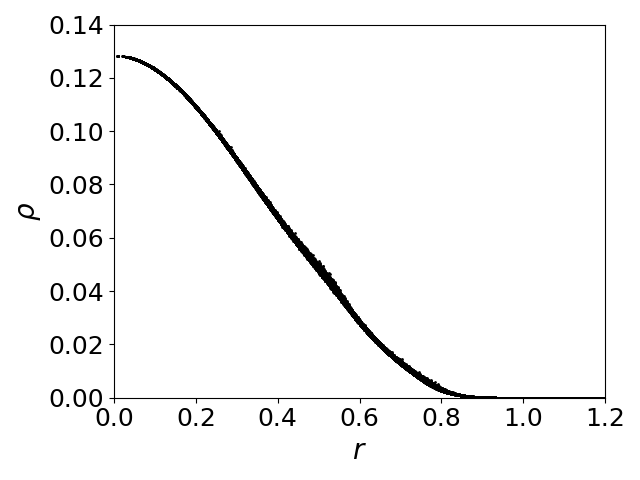}
   \includegraphics[width=0.3\textwidth]{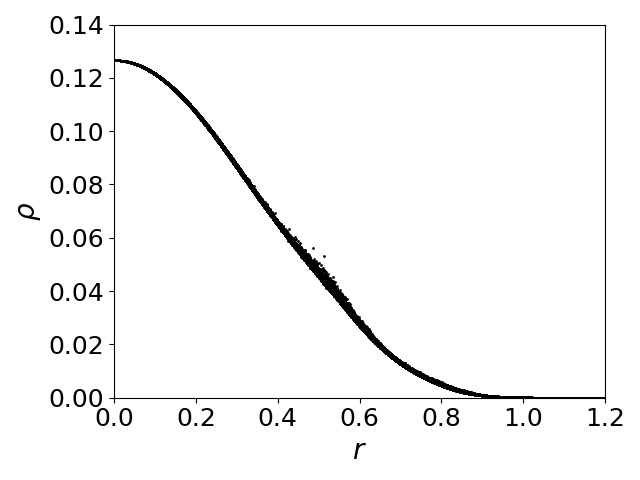}
      \caption{Density colormaps in the $x-y$ plane (top row) and
        density versus radius plots (bottom row) of the star at
        $t/t_{\rm dyn} = 0.035$, $10.5$ and $21$, where $t_{\rm dyn}=1/\sqrt{\rho_c(t=0)}$.  \label{fig:tov}}
 \end{figure*}

We must also compute the associated metric for the TOV equation, which is
\begin{equation}
ds^2 = -e^\nu dt^2 + \left(1 - \frac{2M}{r}\right)^{-1} dr^2 + r^2 d\Omega^2
\end{equation}
The equation for $\nu$ is
\begin{equation}
\frac{d\nu}{dr} = -\left(\frac {2}{P +\mu}\right)\frac{dP}{dr}
\end{equation}
with the boundary condition
\begin{equation}
\exp(\nu) = \left(1-\frac {2M(R)}{R}\right)
\end{equation}

Looking at the Arnold, Dewitt, and Misner (ADM) 3+1 line element for this diagonal metric (\ref{eq:adm metric}), 
we immediately read off the ADM quantities: $\alpha = \exp(\nu/2)$,
$\beta^k = 0$, $\gamma_{rr} = \left(1 - 2M/r\right)^{-1}$,
$\gamma_{\theta\theta} = r^2$, and
$\gamma_{\phi\phi} = r^2 \sin^2\theta$.

\subsection{Evolution of Equilibrium TOV Star}
\label{TOVevol}

We output TOV solution for both the metric and GRHD quantities on a
dense radial grid with significantly higher resolution than
the 3D model to be constructed to set metric and GRHD initial
conditions on our mesh.
We adopt two meshes for simulations presented here, with $10^5$
and $10^6$ mesh generating points, so that a total of $2\times 10^5$
and $2\times 10^6$ mesh generating points are used, respectively, when
the atmosphere is included.  We construct the mesh generating points on a
nearly regular grid for the star, but increase the spacing in a
continuous manner in the atmosphere, which is set to be $10^{-6}$ of
the central density, away from the star, similar to the
initial conditions described in \citet{Prust+19}.  As a result, we
maintain high resolution around the TOV star, but reduce the
resolution in a continuous manner in the atmosphere.

Figure \ref{fig:tov}, shows $x-z$ plane density colormaps and
density-versus-radius plots for the high resolution $N=10^6$ simulation at
$t/t_{\rm dyn}=0.035$ (left), 10.5 (middle), and  21 (right) dynamical
times, $t_{\rm dyn} = 1/\sqrt{\rho_c(t=0)}$; or equivalently 0.1, 30,
and 60 light-crossing times, $t_{\rm LC} = R/c$. Over this timescale, we see that the pressure of the star is roughly
balanced by the ``gravitational'' acceleration (determined by
the appropriate derivatives of the static metric). This balancing is
evident in the stable nature of the star's interior in the upper
projection plots and the relativity static nature of the lower
density-versus-radius plots. We conclude the star is stable.

However, we note that the boundary of the star suffers
from diffusive smearing at later times.  This is evident both in the
projection plots (top row) and profile plots (bottom row) of Figure
\ref{fig:tov}.  This is perhaps unsurprising as the gradients in
density are large near the surface, our spatial reconstruction scheme
is only second order, and we use the HLL Riemann solver.  Both higher
order reconstruction schemes and a less diffusive Riemann solver will
likely alleviate the situation (though see \citealt{2014CQGra..31g5012R}).

\begin{figure}
  \includegraphics[width=0.5\textwidth]{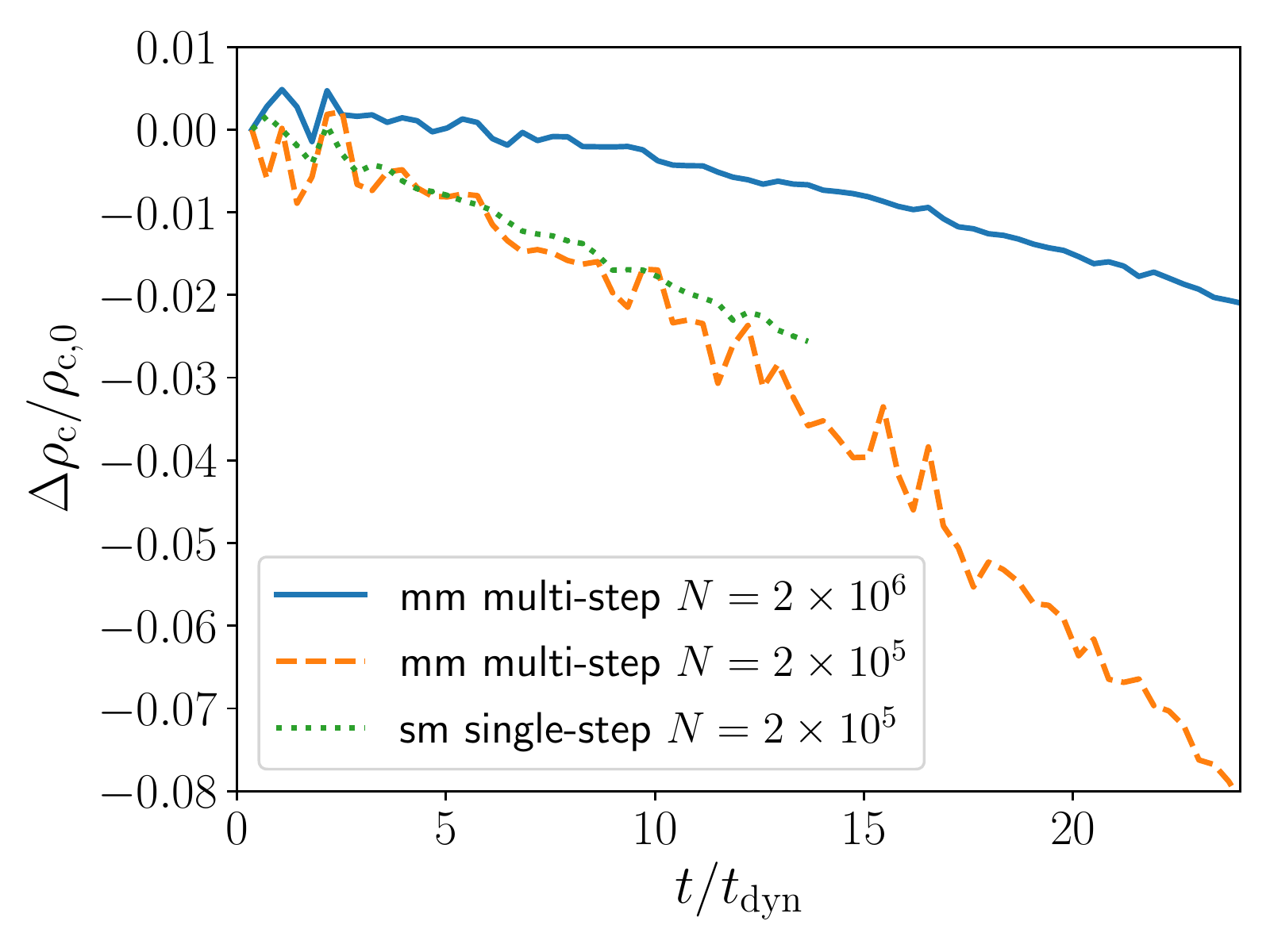}
  \caption{Central density of the non-pressure-depleted TOV star as a
    function of time, at different resolutions.
    \label{fig:central}} 
\end{figure}

In Figure \ref{fig:central}, we plot the central density as a function
of dynamical time and find the central density is accurately
maintained to 2\% for 24 dynamical times at high resolution, confirming the stability of the star.   
However, the central density for the lower resolutions runs suffers a
drift that is about 8\% at 24 dynamical times. In particular,
throughout the bulk of the star the GRHD fields
are smooth and the scheme should converge at second order with
increased resolution. Contrast this with the extremely sharp density
gradient at the surface initially, where we would expect the scheme to
drop to first-order convergent. As the evolution spans many
sound-crossing times, it is possible for first-order convergent
behavior to influence the center of the star. Thus we would expect the
central density drift to converge to zero somewhere between first and 
second-order with increased numerical resolution.

In earlier Newtonian simulations \citep{Chang+20}, the measured
convergence in \changaMM of the L1 norm, $\epsilon$, for weak acoustic 1D
sound waves as a function of linear resolution was found to be
$\epsilon \propto n_x^{-1.74}$. For a 3D distribution of points, we
would therefore expect the convergence to scale at best like $\epsilon
\propto N^{-1.74/3}$, where $N$ is the number of mesh generating
points.   For the mesh generating point ratio between the high and low
resolution simulations of 10, the convergence is then expected to be
$10^{-1.74/3} \approx 0.26$ which agrees with the
central density drift at 24 dynamical times:
$0.02/0.08 \approx 0.25$.

We note that the observed variations in $\rho_c$ are substantially
worse than the results at similar resolution in \citet{IllinoisGRMHD}.
However, it is comparable to the high resolution case of
\citet{GRHYDRO} (65 grid point across the diameter of the star) though we are essentially at double the resolution and
our variations are larger. 
There may be a few reasons for this. 
First, our reconstruction is linear, with error term appearing at
second order. \citet{IllinoisGRMHD} uses the piecewise parabolic
method (PPM)~\citep{1984JCoPh..54..174C}, which is a reconstruction
method accurate to third-order in the numerical error
and thus may be substantially better in static regions with large gradients.
Indeed the observed drift at lower resolutions is comparable to
second-order reconstruction results presented in \citet{Duezetal2005}
on a Cartesian grid. It is well-known that higher-order spatial reconstruction and time integration are crucial to minimizing central density drift in static stars \citep{Duezetal2005}. However, adapting higher order methods to moving-mesh algorithms such as \changaMM remains an open research problem.
Implementation of a less diffusive Riemann
solver, e.g., relativistic HLLC
\citep{2005MNRAS.364..126M,2016ApJS..225...22W} may also alleviate the
situation and we plan to implement this in future work.
  
\subsection{Evolution of Pressure-Depleted TOV Star}
\label{TOVoscilevol}

\begin{figure}
  \includegraphics[width=0.5\textwidth]{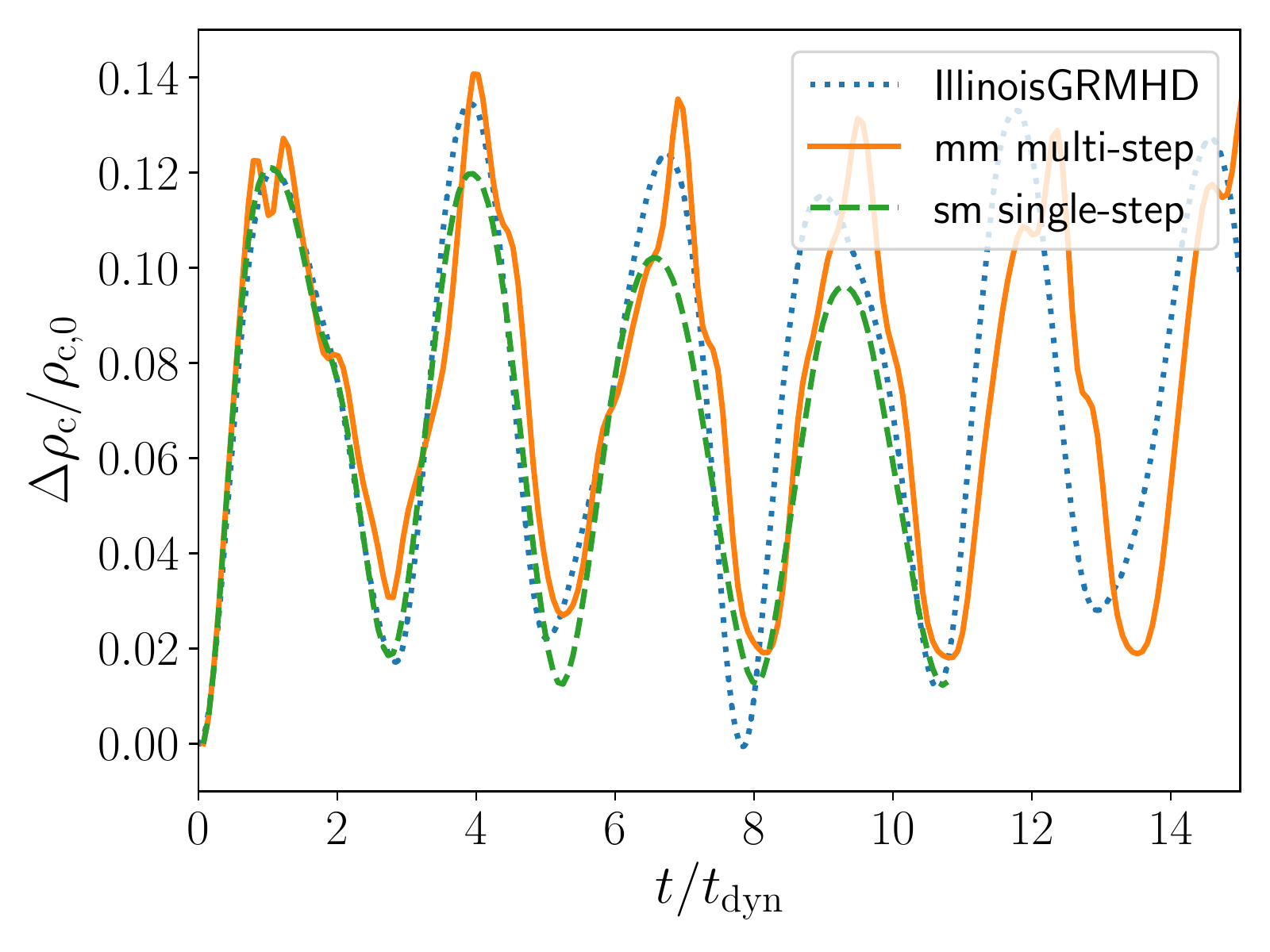}
  \caption{Central density as a function of dynamical time for a star
    whose equilibrium pressure has been reduced by 10\% globally.  We
    show the calculation as run by (1) \IllinoisGRMHD; (2) \changaMM with a
    moving-mesh and multiple timesteps; and (3) \changaMM with a
    static-mesh and a single timestep.
    \label{fig:oscillations}}
\end{figure}

As another test of our code, we use precisely the same TOV star initial
data as the previous section, except we reduce the pressure at every
point by 10\%.  The reduction in the pressure ensures the star is
no longer in hydrostatic equilibrium and will undergo radial
oscillations.  By plotting the central density as a function of time,
we can observe the oscillation of the star about its fundamental mode.
This is shown in Figure \ref{fig:oscillations}.  Here we plot for
$N=10^5$ mesh generating points in the star the case where the mesh is
moving with multiple timesteps (solid line) and the case where the
mesh is static with a single timestep (dashed-line).  For comparison,
we also plot the same calculation run with \IllinoisGRMHD at very high
accuracy ($\approx$126 points across the stellar diameter, as compared
to $\approx$58 points with \changaMM).  The
multiple-timestep approach gives a (slightly) less smooth evolution, but it is notably
faster (by a factor of 10 in this case) and still produces a similar
evolution in terms of frequency and amplitude. Both of these runs 
agree with the evolution of \IllinoisGRMHD.

\begin{figure}
  \includegraphics[width=0.5\textwidth]{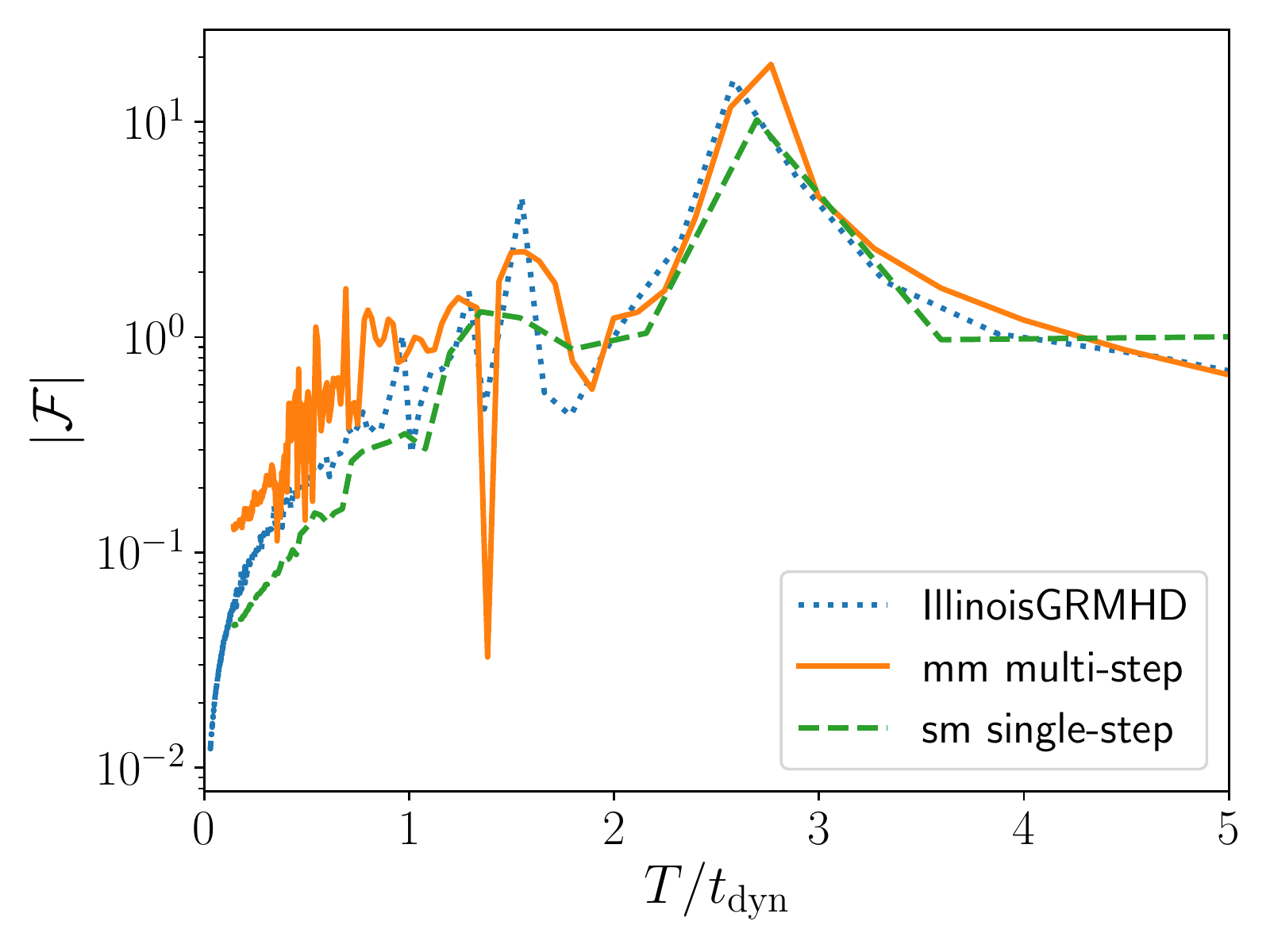}
  \caption{Fourier power (with arbitrary normalization) of the fractional change in central
    density $\Delta\rho_c/\rho_{c,0}$ for a star whose equilibrium
    pressure has been reduced by 10\% globally for a calculation done
    with \changaMM with a moving-mesh and multiple
    timesteps; \changaMM with a static mesh and single timestep; and
    \IllinoisGRMHD. The fundamental mode is the peak near $2.8 t_{\rm dyn}$. 
    \label{fig:fft}}
\end{figure}

A Fourier transform of the fractional variation in central density further confirms qualitative agreement between
\IllinoisGRMHD and \changaMM.  Here we apply a Gaussian window
function of the form $\exp(-t^2/t_0^2)$ to and take a Fourier
power of $\Delta\rho_c$, where $t_0 = 10 t_{\rm dyn}$. Changing
the value of $t_0$ between 5 and 20 $t_{\rm dyn}$ do not change our
results significantly.   In Figure \ref{fig:fft}, we plot the
Fourier power, $|\mathcal{F}|$, as a function of the period, $T$, in
units of $t_{\rm dyn}$ with arbitarily normalization. 
The crucial feature of this plot is that three runs agree on the
position of the fundamental peak at $T\approx 2.8 t_{\rm dyn}$, though
it is less clear in the \changaMM run with the static mesh and single
timestep due to the shorter time series.

\section{Conclusions and Future Work}\label{sec:conclusions}

We have reviewed the structure and implementation of a moving-mesh
general relativistic hydrodynamics solver for static spacetimes in
\changaMM. Taking advantage that any flux-conservative equation can be
solved on a moving, unstructured Voronoi mesh, we briefly describe the
general algorithm to solve the GRHD equations. We then elaborate on a
number of technical points including implementation of the van Leer
method of second-order time integration, strategy for solving the Riemann
problem across moving faces, transformation of primitives to/from
conservatives variables, validation and correction of
hydrodynamic variables, and mapping a spherically symmetric
metric to a Voronoi grid.  

We then apply our code to the analytic solution of a TOV star and show
that our algorithm integrates the star stably as expected.  In
particular, we show that at high resolution ($10^6$ mesh generating
points inside the star) that the star is stable and the central
density varies by at most 2\% over 24 dynamical times.  At lower
resolution ($10^5$ mesh generating points inside the star), the
central density suffers a systematic drift toward lower density
similar to that observed by \citet{Duezetal2005} on Eulerian
(Cartesian) grids at low resolution with second order spatial
reconstruction.  Because our current MM method is limited to
second-order reconstruction, we find that our results are consistent
with that of \citet{Duezetal2005}. Moreover, we also find that we can
mitigate this effect in large part by going to higher
resolution. Finally we demonstrate that when evolving the same star,
but with its initial pressure depleted by 10\%, we recover the same
fundamental frequency of ensuing oscillations.

This paper is the first in a series that will describe the eventual
implementation of a moving-mesh GRHD code on \textit{dynamical}
spacetimes.  While the code can already be used to study flows around black holes and neutron stars, this is not our primary interest.  Instead we aim to study the mergers of compact binaries with matter.  Toward that end a number of future improvements are planned and will be described in future work.  These are listed in order of priority:
\begin{enumerate}
   \item \textit{Incorporation of a full dynamical spacetime solver.}
     While there are many applications of a moving-mesh GRHD solver in
     static spacetimes, our main interest is adapting our moving-mesh
     GRHD to NS-NS and NS-BH mergers.  As such, we are incorporating
     the dynamical spacetime solver \SENRNRPy in \changaMM.  Here the
     major challenge is to couple the \changaMM and \SENRNRPy codes,
     and get them to pass information between one another. We note
     that although \SENRNRPy is currently geared toward solving the 
     binary BH problem (for which $\Tmunu=0$), all $\Tmunu$
     source terms were recently added to Einstein's equations within
     \SENRNRPy and validated\footnote{\url{https://nbviewer.jupyter.org/github/zachetienne/nrpytutorial/blob/master/Tutorial-Start_to_Finish-BSSNCurvilinear-Neutron_Star-Hydro_without_Hydro.ipynb};
       download here: \url{https://github.com/zachetienne/nrpytutorial}} for the case of a TOV star with a fixed
     stress-energy tensor but an evolving spacetime metric (the so-called
``hydro without hydro'' test of numerical
relativity;~\citealt{Baumgarte:1999tz}).
   \item \textit{Implementation of piecewise-polytropic and tabulated
     equations of state.} As part of our strategy to create realistic
     simulations of NS-NS and NS-BH mergers, we plan to implement
     realistic equations of state into \changaMM with the GRHD solver.
     In this regard, we plan to both implement piecewise-polytropic
     and tabulated equations of state in \changaMM. In fact, a nuclear
     equation of state has already been implemented
     \citep{2017PhRvC..96f5802S} in the Newtonian version of
     \changaMM.   The framework for implementation of generic
     equations of 
state is general and can be used (in the Newtonian case) for both the
moving-mesh solver and the SPH solver (if required) of \changaMM
without modification.  Here we plan to modify the framework to support
the GRHD solver.  
   \item \textit{Implementation of GRMHD.} Magnetic fields play an
     important dynamical role in compact binary mergers with matter.
     Toward that end, we plan on implementing GRMHD in \changaMM.  To
     do this, we will need to modify the equation of hydrodynamics to
     include magnetic fields and include an evolution equation for the
     magnetic field.  In moving, unstructured meshes, two different
     schemes have emerged for evolving the magnetic field while
     maintaining the divergence-free condition, $\partial_iB^i = 0$,
     where $B$ is the magnetic field.  These schemes are the
     divergence cleaning/diffusion methods, e.g., Dedner scheme
     \citep{2002JCoPh.175..645D,2011MNRAS.418.1392P} and vector
     potential methods \citep{2016MNRAS.463..477M}.  We have
     implemented an MHD scheme in \changaMM based on the vector
     potential scheme of \citet{2016MNRAS.463..477M} and plan to adapt
     the scheme to the general relativistic case.  As this method is
     similar to the currently implemented method in \IllinoisGRMHD, we
     do not anticipate major technical challenges.
   \item \textit{Implementation of neutrino physics/radiation.}
     Neutrinos affect the outflows and eventual r-process yields of
     NS-NS and NS-BH mergers, and can act as energizers of the
     outflows as well as changing the $Y_e$ of the resulting
     outflowing material to change both the total mass of the outflow
     and its composition.  To capture this physics, we plan to adapt
     our recent time-dependent radiative transfer algorithm
     \citep{Chang+20} to include neutrino radiation.  In particular,
     if we ignore metric effects on the radiation, e.g., straight-line
     propagation, the adaption of our current radiation methods would
     require the expansion from one photon species to the number of
     neutrino species and the addition of an evolution equation and
     source terms for the electron fraction \citep[see for
       instance][]{2010ApJS..189..104M,2019ApJS..241....7S}.  
\end{enumerate}

\section*{Acknowledgments}

We acknowledge Charles Gammie, Andrew MacFadyen, Philipp Moesta, Kenta Hotokezaka, and Chris White for useful discussions.  We thank Paul Duffell and Francois Foucart for crucial insights on the computation of the Riemann flux on unstructured grids. We also thank Scott Noble for providing a thread-safe version of his conservative-to-primitive solver.  We thank the anonymous reviewer for helpful comments and criticisms.
PC is supported by the NASA ATP program through NASA grant NNH17ZDA001N-ATP. PC acknowledges the support and hospitality of the Center for Computational Astrophysics at the Flatiron Institute where much of this work was carried out.  
ZBE gratefully acknowledges the NSF for financial support from awards
OIA-1458952 and PHY-1806596; and NASA for financial support from
awards ISFM-80NSSC18K0538 and TCAN-80NSSC18K1488.  
The authors acknowledge the Texas Advanced Computing Center (TACC) at The University of Texas at Austin for providing HPC resources that have contributed to the research results reported within this paper, \url{http://www.tacc.utexas.edu}.
Computations were performed on the Niagara supercomputer at the SciNet HPC Consortium. SciNet is funded by: the Canada Foundation for Innovation; the Government of Ontario; Ontario Research Fund - Research Excellence; and the University of Toronto. \IllinoisGRMHD
simulations were completed on the Thorny Flat HPC System at West
Virginia University (WVU), which was funded in part by the
National Science Foundation (NSF) Major Research Instrumentation
Program (MRI) Award \#1726534, and WVU. We also use the \texttt{yt}
software platform for the analysis of the data and generation of plots
in this work \citep{yt}. The Flatiron Institute is supported by the
Simons Foundation.






\bibliographystyle{mnras}
\bibliography{references}




\bsp	
\label{lastpage}
\end{document}